\documentclass[]{aastex631}

\submitjournal{AJ}

\shorttitle{Post-Red Supergiants in the LMC}
\shortauthors{Humphreys, Jones, \& Martin}

\begin{document}

\title{Yellow Supergiants and Post-Red Supergiant Evolution in the Large Magellanic Cloud}

\correspondingauthor{Roberta Humphreys}
\email{roberta@umn.edu}

\author{Roberta M. Humphreys}
\affiliation{Minnesota Institute for Astrophysics,  
University of Minnesota,
Minneapolis, MN 55455, USA}

\author{Terry J. Jones}
\affiliation{Minnesota Institute for Astrophysics,
University of Minnesota,
Minneapolis, MN 55455, USA}

\author{John C. Martin}
\affiliation{University of Illinois, 
Springfield, IL}

\begin{abstract}
	The empirical evidence for an upper mass limit for the red supergiant (RSG)  progenitors of the Type II-P SNe at about 18 Msun, raises questions about the fate of the most luminous, most massive RSGs.  These stars may evolve back to warmer temperatures to end their lives as hotter stars or collapse directly to black holes. The yellow hypergiants, many with extensive circumstellar dust and high mass loss, are excellent candidates for post-RSG evolution. We have identified six high luminosity yellow supergiants (YSGs) in the LMC with circumstellar dust including two of the FYPS \citep{Dorn22} . We discuss their SEDs, mass lost and mass loss rates.  Together with three additional FYPS, these nine stars are about 1/3 of the YSGs above 10$^{5}$ L$_{\odot}$.  We conclude that the high luminosity YSGs with surface pulsations and circumstellar dust, distinct from other YSGs, are candidates for post-RSG evolution in the LMC.   
\end{abstract}

\keywords{galaxies: individual(LMC) stars: supergiants stars: massive stars: evolution}

\section{Introduction} \label{sec:intro}
Decades of observations of the most luminous stars in the Milky Way and nearby galaxies have revealed a complex population of evolved massive stars whose evolutionary histories and eventual fate depend not only their initial mass but rotation, possible binarity, etc and most importantly mass loss and their mass loss histories. The numerous surveys for supernovae have increased the variety of types of terminal events and other eruptive phenomena, {such as the non-terminal giant eruptions like eta Carinae and the ``supernova impostors'', Luminous Blue Variables (LBVs), and high mass loss episodes from the yellow and red hypergiants.} Consequently, the diversity of possible progenitors have raised questions about the end-state of the different groups of evolved massive stars.

\vspace{2mm}

Red supergiants(RSGs) were long considered the end-product of stellar evolution for stars ~$\approx$ 9 to 40 M$_{\odot}$. Thus the majority of massive stars were expected to end their lives on the red side of the HR Diagram as Type II-P or Type II-L supernovae. But SN 1987A occurred on what was considered, at least at that time,  the `wrong side of the HR Diagram' {\citep{Arnett}}. More recently, \citet{Smartt,Smartt15} identified what is now known as the ``red supergiant problem'', the lack of Type II RSG progenitors above 18 M$_{\odot}$; {confirmed statistically in a recent independent survey by \citet{Rodriguez}.}  The more massive RSGs, 20 to 40 M$_{\odot}$  could possibly collapse directly to the black hole {\citep{Sukhbold16,Sukhbold18}} or evolve back to warmer temperatures before the terminal explosion. As post-RSGs, they would evolve back across the HR diagram, through the region of the yellow supergiants (YSGs), a relatively short-lived state, and thus a rather sparsely populated region of the HR Diagram. But how to tell them apart. Are they evolving from blue to red as expected or from red to blue in a more evolved state having been RSGs?
 
\vspace{2mm} 

The red supergiant stage is a well-known enhanced mass loss state \citep{Mauron}. Stellar structure models show that if the stars shed sufficient mass as RSGs they can evolve to warmer temperatures. 
The most luminous RSGs luminosities $\ge$ 10$^{5}$ L$_{\odot}$ have mass loss rates of $\>$  
10$^{-5}$ with more extreme stars like VY CMa and NML Cyg reaching 10$^{-4}$ M$_{\odot}$ yr$^{-1}$. On this second passage across the HR Diagram, these post-RSGs having shed additional mass, will be more subject to surface instabilites driven by their warmer temperatures and increased rotation, and thus be identified by their continued mass loss and circumstellar ejecta. 

\vspace{2mm} 

Very few post-red supergiants or candidates are known. In the Milky Way, the best example is IRC~+10420 \citep{Jones,Oud96} with its complex circumstellar ejecta, high mass loss rate and history of episodic mass loss \citep{RMH97, RMH02, Tiffany, Oud98, Shenoy}. The `Fried Egg Nebula' (IRAS ), a post-RSG candidate with extended nebulosity, was recently confirmed to have multiple outflows \citep{Koumpia}. The peculiar yellow hypergiant \footnote{We use the term yellow or warm hypergiant for the most luminous YSGs near the upper luminosity boundary in the HR Diagram.}  Var A in M33 \citep{RMH06}  experienced a high mass loss episode with an apparent transit to lower temperatures that lasted decades, and is an excellent example of a star in probable post-RSG evolution. These examples all have high luminosities, high mass loss, and spectra with strong  emission lines due to mass loss and their stellar winds.  Consequently, \citet{RMH13} identified several warm hypergiants in M31 and M33 with dusty spectral energy distributions (SEDs) and emission line spectra with strong wind lines. Adopting these  criteria, \citet{Gordon} found that 30 --40\% of the YSGs in these two galaxies are candidates for post-RSG evolution. 

\vspace{2mm} 

Using the Transiting Exoplanet Survey Satellite (TESS), \citet{Dorn20} recently  identified a subset of A and F-type supergiants in the LMC with rapid surface pulsations. Pulsations are not expected in the YSGs on their first crossing of the HR Diagram, but after the onset of He burning as a red supergiant, their interior structure is altered and the post-RSGs may pulsate. These Fast Yellow Pulsating Supergiants (FYPS) have luminosites above 10$^{5}$ L$_{\odot}$ corresponding to stars with initial masses above 18-20 M$_{\odot}$. \citet{Dorn22} thus argue that the FYPS are consistent with post-red supergiant evolution.   

\vspace{2mm} 

As part of a larger study of the massive star populations in the LMC, \citet{Martin}, identified over 80 yellow supergiants with spectral types and multi-wavelength photometry. Here we describe the YSG survey and report on six post-red supergiant candidates.   In the next section we describe the star selection. The post-RSG candidates, their SEDs, the DUSTY model for their circumstellar dust, and mass loss rates are discussed  in \S {3}. The properties of these stars and their positions on the HR Diagram are summarized in the last section. 

\section{Star Selection}  

Due to the high foreground contamination in the temperature range of the yellow supergiants (4500 to 8000K), we have limited our selection of LMC YSGs to those with spectral classification types from F0 to the early K-type. The warmer A-type supergiants and the red supergiants, spectral types K5 and later, are not included in this study.  

\vspace{2mm}

Many of our member stars were selected from the early surveys of the hot and luminous stars in the Magellanic Clouds including photographic spectral classification and photoelectric photometry; \citet{Feast}, \citet{Ardeberg}, \citet{Brunet}, \citet{SOI} and the compilation by \citet{Rousseau}, {plus occasional single stars from other early surveys, such as \citet{West}, \citet{MG}, and \citet{SandPhilip}.}  Since then, \citet{MO} and \citet{Neugent} have used CCD imaging  and radial velocities to select probable yellow and red supergiant members. Additional stars with spectral types were added from their lists and two more recent searches for members \citep{GF15,Davies}. {All of the candidate member stars were checked against the Gaia DR3 Catalog. Only one star was removed with a parallax and proper motion inconsistent with LMC membership.}
 
\vspace{2mm} 

This yielded a list of 84 YSGs  with spectral types and visible and near-infrared photometry. We derived their visual interstellar extinction  (A$_{v}$) from their colors and spectral types;  their visual luminosities at our adopted distance modulus of 18.5 mag for the LMC, and the corresponding bolometric luminosities.  We determine the bolometric luminosity from the K -band photometry following the description in \citet{Neugent}. Their luminosities, however,  depended on an adopted mean constant color excess for the LMC of 0.13 mag.  We find a larger mean  E(B-V) of 0.22  for the YSGs. Consequenty, we used the color excess for each individual star to derive the K band extinction and the adopted temperature from the spectral type for the K band bolometric correction. The resulting luminosities average 0.14  higher in Log L/L$_{\odot}$  than those in \citet{Neugent} and \citet{Dorn22}.

\section{Circumstellar Dust, Mass Loss and the Post Red Supergiant Candidates}

The Galactic post-RSG candidates like IRC~+10420 are distinguished by their high luminosities and high mass loss with significant excess radiation longwards of 2 microns due to circumstellar dust. Their spectra have strong emission lines of Hydrogen, Ca II triplet and [Ca II] with P Cygni profiles from their stellar winds. We used these criteria to identify post-RSGs in M31 and M33 \citep{Gordon}.
 
\vspace{2mm} 

The spectroscopic criteria, however, are a problem for most of the YSGs in the LMC. The majority were identified in photographic surveys in the 1970's. The  spectra were obtained in the photographic blue spectral region and do not include the longer wavelengths with the strong emission lines including H$\alpha$. We searched the published papers for notes on the spectra and comments on emission lines. A few stars were found with Hydrogen emission, mostly due to nebulosity. For example, our survey of the literature for F-type supergiants in the LMC surrepticiously picked up two known LBVs in their optically thick wind, high mass loss state, and of course with emission lines. They are not included here, but will be discussed in a later paper.  A few stars with Hydogen emission are described briefly in the subsection below. Consequently, in this paper we rely on the presence of circumstellar dust at the long wavelengths in the their SEDs to identify candidates. 

\vspace{2mm} 

With the numerous mid-infrared surveys including the {\it Spitzer} IRAC SAGE survey \citep{SAGE}, WISE \citep{Wright} and Akari \citep{Akari}, we can easily search their  on-line databases for excess infrared radiation in the  YSGs.  All 84 YSGs were checked for excess radiation  in the 5 to 24 $\mu$m region. Six stars were identified with circumstellar dust in their SEDs including two FYPS pulsators \citep{Dorn22}.  It is not known if the four other stars are FYPS  pulsators. They do not have TESS data at the two-minute cadence used in the \citet{Dorn22} study.  The published photometry for these stars and the five FYPS F-type supergiants is\footnote{HD 269651 is listed as F0 I in \citet{Dorn22}, but the description of its spectrum and published photometry supports the A2 Ia0 type \citep{Ardeberg}.} summarized in Tables 1A and 1B.

We note that three of the four high luminosity F-type supergiants classified by \citet{Keenan} as luminosity class 0 have circumstellar dust, and two are known pulsators. Excess mid-infrared radiation in these three stars was also reported by \citet{Kour}. The fourth star, HD 271182 (F8 0) has no circumstellar excess nor is it a known FYPS  pulsator.

\begin{deluxetable*}{lllllllllll}
\tablewidth{0pt}
\tabletypesize{\small}
\tablenum{1A}  
\tablecaption{Visual and Near-Infrared Photometry}
\tablehead{
\colhead{Star} & 
\colhead{Sp Type}&
\colhead{V}  &
\colhead{B-V} &
\colhead{U-B}&
\colhead{R} &
\colhead{I} &
\colhead{J} &
\colhead{H} &
\colhead{K} & 
\colhead{Comment} 
}
\startdata
HD 268687 & F6 Ia & 10.65 & 0.47 & 0.21 & 10.62\tablenotemark{b} & 10.50\tablenotemark{b} & 9.69 & 9.45 & 9.36 & FYPS, no CS excess \\ 
HD 268757 & G8 0 & 10.09&  1.55	& 1.29     & 9.21\tablenotemark{a} & 8.52\tablenotemark{a}  & 8.02 & 7.64	& 7.45  & R59, CS excess  \\  
HD 269154 &  F6 Ia & 10.50 & 0.50 &  -0.06 & \nodata  & \nodata   & 9.56  & 9.34 & 9.19 & CS excess  \\ 
CD-69 310 & F2 I & 10.70 & 0.26 & 0.17      & \nodata & \nodata & 9.98 & 9.84& 9.78 &  FYPS, small excess? \\ 
Sk -69 148 & K0 I & 10.93 & 1.61  & \nodata &  10.23\tablenotemark{b} & \nodata  &  8.97& 8.67	& 8.38	 & CS excess \\  
HD 269723   & G4 0 & 9.91 & 1.15 & 0.60     & \nodata & \nodata & 8.23  & 7.88  & 7.69  & FYPS, R117, CS excess \\  
HD 269840 &  F3 Ia & 10.32 & 0.42 &  0.21   & \nodata &  9.76\tablenotemark{b} &  9.31 & 9.11 & 8.99  &  FYPS no CS excess  \\
MG73 59 & K0 I &  10.68	& 1.52	& \nodata  & 9.87\tablenotemark{b} & \nodata  & 8.48  & 8.04 & 7.81 & CS excess  \\ 
HD 269953  &  G0 O & 9.93 & 0.87& 0.62      & \nodata & \nodata &  8.59	& 8.33	& 8.02  & FYPS, R150, CS excess   
\enddata
\tablenotetext{a}{R, I Johnson}
\tablenotetext{b}{R, I Cousins-Kron}
\end{deluxetable*}

\begin{deluxetable*}{lllllllllllllll}
\tabletypesize{\scriptsize}
\tablenum{1B}  
\tablecaption{Multi-Wavelength Mid-Infrared Photometry }
\tablewidth{0pt}
\tablehead{
\colhead{Star} & 
\colhead{3.6$\mu$m}\tablenotemark{c} &
\colhead{4.5$\mu$m}\tablenotemark{c} &
\colhead{5.8$\mu$m}\tablenotemark{c} &
\colhead{8$\mu$m}\tablenotemark{c}  &
\colhead{24$\mu$m}\tablenotemark{d} &  
\colhead{3.4$\mu$m}\tablenotemark{e} &
\colhead{4.6$\mu$m}\tablenotemark{e} &
\colhead{12$\mu$m}\tablenotemark{e} &
\colhead{22$\mu$m}\tablenotemark{e} &  
\colhead{7.1$\mu$m}\tablenotemark{f}&  
\colhead{10.5$\mu$m}\tablenotemark{f} & 
\colhead{15.6$\mu$m}\tablenotemark{f}  & 
\colhead{18.4$\mu$m}\tablenotemark{f}  & 
\colhead{22.9$\mu$m}\tablenotemark{f}  
}
\startdata
HD 268687 & 9.30 & 9.17 & 9.09 & 8.99 & 8.43  & 9.22 & 9.15 & 8.96 & 9.05 &  \nodata & \nodata  & \nodata  & \nodata  &  \nodata     \\ 
HD 268757  & 7.20   & 7.55 & 7.25 & 7.29 & 5.45  & 7.07& 7.09	&6.80	&5.39	& \nodata  & \nodata  & \nodata  &  \nodata & \nodata   \\  
HD 269154  &  9.00&  8.88 & 8.83 & 8.81 &   7.92  &  8.94  & 8.86  & 8.65 &  7.36 & 8.84  & 3.67 & 1.30   & \nodata  & 0.99  \\ 
CD-69 310  & 9.63 & 9.57 & 9.55 & 9.54    & 9.56  &  9.66 & 9.63 & 9.45 &  7.78 & 4.59  &  1.91 & 0.52  & \nodata  & 0.21  \\ 
Sk -69 148 & 8.20  & 7.90 & 7.70 &7.43 & 5.98  & 8.13 & 7.89 & 7.04 &  6.03 &  30.9 &  20.7&  7.97 & \nodata & 4.40   \\  
HD 269723  & 7.45 & 7.32 &  7.25& 7.24& \nodata  & 7.34& 7.27	& 6.21 &1.58& \nodata & \nodata  & \nodata  & \nodata & \nodata  \\  
HD 269840 & 8.85 & 8.78 & 8.76 & 8.72 & 8.16 & 8.84 & 8.78 & 8.58 & 8.33 & 18.4  & 3.70  & \nodata   &  \nodata  & \nodata  \\
MG73 59   & 7.61 & 7.62 &  7.21	& 6.61	&  3.79 &  7.38	& 7.31	& 5.54	& 3.77	&  58.5 & 69.7 &  32.9 & \nodata  & 29.7  \\ 
HD 269953   & 7.28 & 6.70 & 6.22 & 5.16 & 1.95  &  7.27& 6.64  & 4.06 & 1.21 & 161  &  318  & 156  & 218& 169    \\ 
\enddata
\tablenotetext{c}{{\it Spitzer}/IRAC}
\tablenotetext{d}{{\it Spitzer}/MIPS}
\tablenotetext{e}{WISE}
\tablenotetext{f}{Akari in units of 10$^{-15}$ Watts m$^{-2}$}   
\end{deluxetable*}

\vspace{2mm} 

The SEDs for the six YSGs with circumstellar dust are shown in Figure 1.  
Although the YSGs or F-type supergiants are the focus of this work, we examined the infrared fluxes for the high luminosity A-type supergiants and FYPS in \citet{Dorn22} for CS dust and found only one marginal case, HD 269661 (A0 Ia0e) which is also an emission line star. Two of the less luminous pulsators (log L/L$_{\odot}$ $<$ 5.0) show evidence for weak circumstellar dust, and none of the non-pulsators have a long wavelength excess due to dust.

\begin{figure}
\plotone{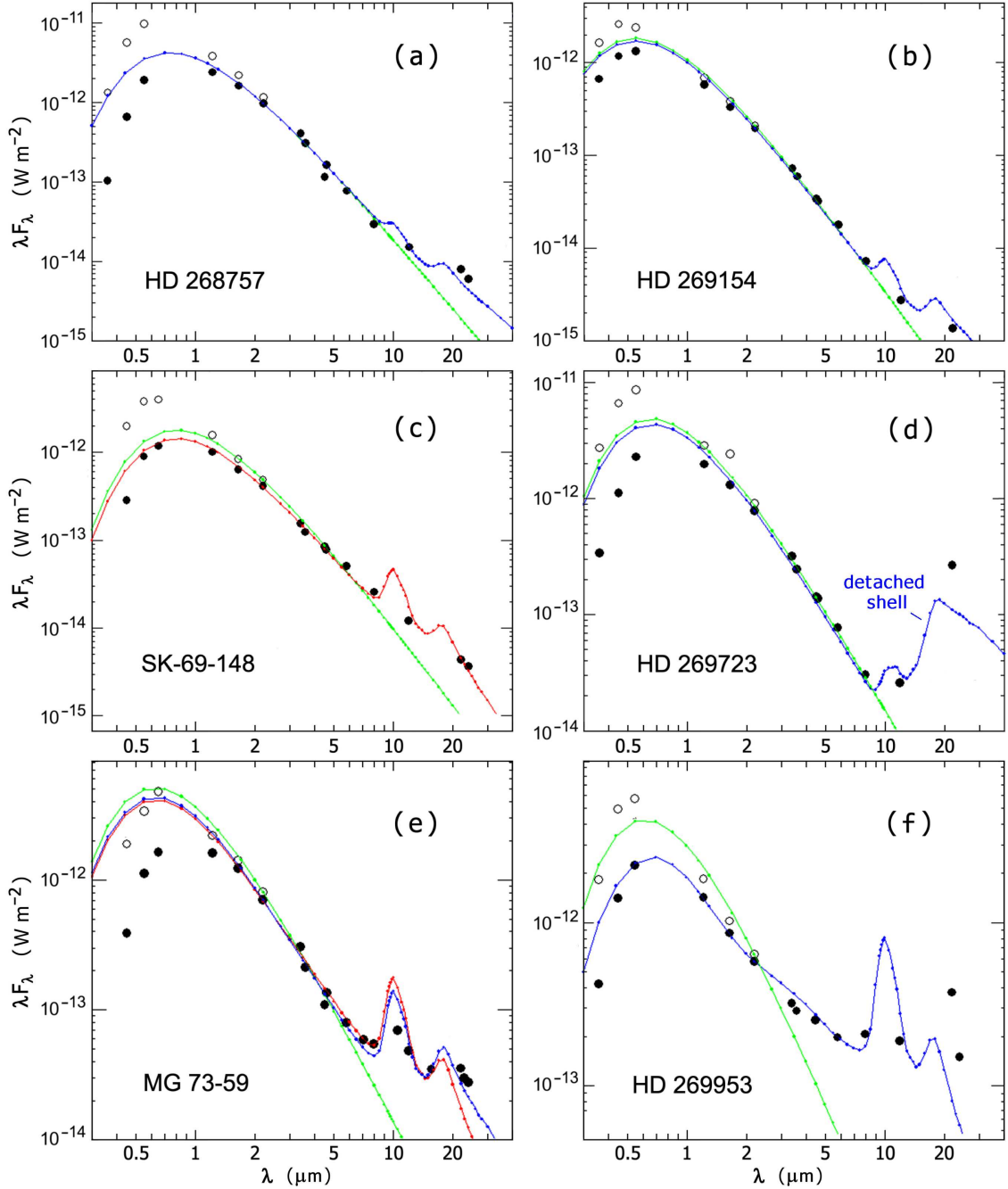}
\caption{The SEDs for the six YSGs with circumstellar excesses. { The observed fluxes are shown as filled circles. The extinction corrected fluxes for the visual and near-infrared are  open circles. The Planck curve fits are shown in green  and the Dusty models in blue and red. Two possible fits are shown for MG 73-59;  red for a constant mass loss rate (n $= 2$) and blue for n $= 1.5$.  The detached shell model for HD 269723 is marked.} The derived parameters for the Dusty models are in Table 2.}
\end{figure}  

\vspace{2mm}

The SEDs show the visual (UBV) and near-infrared (JHK) fluxes corrected for interstellar exinction and the mid-infrared photometry from the Spitzer, Akari and WISE surveys. A Planck curve, corresponding to the star's temperature based on its  spectral type fit through the JHK points and the fit from the radiative transfer code Dusty \citep{Dusty} are shown. We use Dusty to estimate the mass loss rate and the density distribution of the dust and gas. The input includes the optical parameters of the grains; their chemistry, size, and the dust condensation temperature which determines the condensation radius, r$_{1}$. We use the ``cool'' circumstellar silicates \citep{Ossen}, and assume that they follow the MRN size distribution, n(a)$\,\propto$ a$^{-3.5}$da \citep{Mathis}  with a$_{min}$ = 0.005\,$\mu$m and a$_{max}$ = 0.5\,$\mu$m.
We ran a series of models with the adopted stellar temperatures and a fixed shell extent, 1000 r$_{1}$. 

\vspace{2mm} 

Dust condensation temperatures for silicate-rich dust shells range from 700--1000K. For this study we adopt 1000K for the dust condensation temperature of the inner shell, T$_{in}$, \citep{Suh}, and vary the optical depth  $\tau_{v}$ of the circumstellar {dust} from 0.1 to 1.0. In DUSTY the density distribution function is modeled as a power law, $\rho$(r) $\propto$ r$^{-n}$, with $n=2$ for a constant mass loss rate.  A lower index indicates a higher mass-loss rate in the past and a decline over time. {In our models, we allow the density distribution to vary and  find that a lower index was required for some of the YSGs for the best fit to their long wavelength fluxes.  Since we are interested in estimates of the dust shell optical depth and mass-loss rates, the DUSTY model fits are based on the flux longward of 2 $\mu$m.  The model results clearly departed from the 2-10 $\mu$m data when the optical depth was changed by more than $\pm20$\% and the power law index was changed by $\pm$ 0.1. Most of the stars are fit with a constant mass loss rate ($n=2$). The best-fit model was then selected by-eye.  MG73 59 is an example where two different distributions give nearly equally good fits to the 10 and 20$\mu$m emission peaks}.

\vspace{2mm} 

 The parameters for the best DUSTY models are summarized in Table 2. The dust mass loss rate was derived using {the familiar equation $\dot M(t)$ $=$ 4$\pi$r$^{2}$$\rho$(r)\,v$_{exp}$ for a constant mass loss rate, $n=2$. For a non-constant rate we used the equation derived in the Appendix in \citet{RMH2020}.}  Both formulations require  the expansion or outflow velocity. Since we lack spectra showing the emission line profiles for these YSGs,  we rely on measurements of other yellow hypergiants.    IRC ~+10420 has an outflow velocity of about 75 km s$^{-1}$ from the split emission line profiles and the post-RSG candidates in M31 and M33 have outflow velocities of 100 to 200 km s$^{-1}$. Most of these stars have earlier spectral types and warmer temperatures than many of our LMC candidates. We therefore adopt 100  km s$^{-1}$ for the earlier types and 75  km s$^{-1}$ for the cooler YSGs,  G8 to K0. For the total mass loss rate in Table 2, we adopted a gas to dust ratio of 200 for the LMC \citep{Decin,Mauron}.

\begin{deluxetable}{cclllll}
\tablewidth{0 pt}
\tabletypesize{\small}
\tablenum{2} 
\tablecaption{Parameters for LMC YSGs with Circumstellar Dust}
\tablehead{
\colhead{Star} & 
\colhead{Sp Type/T$_{eff}$K} &
\colhead{log L/L$_{\odot}$} &
\colhead{r$^{-n}$} & 
\colhead{$\tau_{v}$} &
\colhead{r$_{1}$ (AU)} & 	
\colhead{M$_{\odot}$ yr$^{-1}$}   
}
\startdata
	HD 268757 & G8 0/5000 &  5.7 & 1.3   &	0.01 &  130  &  6.6 $\times 10^{-6}$   \\  
	HD 269154 & F6 Ia/6800 & 5.4  &  1.3  & 0.006   & 114  & 4.6 $\times 10^{-6}$  \\ 
	Sk -69 148 & K0 I/4600  & 5.6  & 2  & 0.1   & 110  &  2.5 $\times 10^{-6}$   \\  
	HD 269723  & G4 0/5600  & 5.7  & 2  & 0.1   & 8000\tablenotemark{a}  &  6.1 $\times 10^{-5}$  \\  
	MG73 59    & K0 I/4600  & 5.6  & 2  & 0.1   & 134  & 3.1 $\times 10^{-6}$   \\ 
	   "       &   "	   &  " & 1.5 & 0.05 & 132  & 6.4 $\times 10^{-6}$ \\   
	HD 269953  & G0 0/6000  & 5.7  & 2  & 0.7   & 168   &   3.6 $\times 10^{-5}$   
\enddata
	\tablenotetext{a}{Detached shell presumably from a prior RSG stage. Assuming it is from a high mass lost state as a RSG, we used 25 km s$^{-1}$ for the expansion velocity. See text.} 
\end{deluxetable}

\vspace{2mm}

We chose to fit HD 269723 with a detached shell. Note that its SED shows no signature of thermal dust emission above the  reddened photosphere from the near-IR out to about 8 $\mu$um, but beyond 10 $\mu$m there is a clear excess in the WISE 12 and 22 $\mu$m bands. If HD 269723 has an older, now detached, shell it developed in a previous epoch of mass loss, we can fit the SED fairly well using a shell with an inner temperature of 150K and a modest optical depth of tau=0.1 at 
0.55 $\mu$m.   Given the relatively large WISE beam, the long wavelength excess could be due to contamination from warm dust associated with an HII region that may be contributing to the total flux in the beam. We note that 150K is relatively warm for HII region dust, and the lack of PAH emission at 8 $\mu$m, typically seen in the SED of HII regions, suggests the SED is not contaminated by another source in the beam.

\vspace{2mm}

The mass loss rates range from 3  $\times$ 10$^{-6}$ to 6 $\times$ 10$^{-5}$ M$_{\odot} yr^{-1}$, comparable to the rates for the YSGs in M31 and M33. The highest rate is for the detached shell model for HD 269723 with an inner shell radius of 8000 AU. If we assume that this dusty shell is a remnant from a high mass loss state as a RSG, with an outflow velocity of $\approx$ 25 km s$^{-1}$ for a RSG, then it corresponds to a cessation of mass loss about 1500 yrs ago, consistent with the timescale for a  40 M$_{\odot}$ star that has just left the RSG stage  on a  transit  to warmer temperatures \citep{Eggen}.  We can estimate the dust  mass\footnote{The dust mass is given by  \[  M_{dust} = \frac{4D^{2} {\rho} {\lambda}F_{\lambda} }
	             { 3({\lambda} Q_{\lambda}/a) B_{\lambda}(T) }   
		                \] } in the shell  with our  assumed grain properties and adopting   the peak flux at 22 $\mu$m with a grain temperature of 150K. The total mass, gas + dust,  is 10$^{-2}$ M$_{\odot}$.  We consider this a lower limit since we do not know how much colder dust there may be emitting beyond 20 $\mu$m. HD 269723's mass loss rate and mass lost are comparable to those measured in the most luminous RSGs \citep{RMH2020,Gordon} and the mass shed in episodic events \citep{RHTJJ}.

\subsection{Emission Line Stars}

The presence of strong emission lines with P Cygni profiles are indicators of stellar winds, and possible recent enhanced mass loss episodes similar to the yellow hypergiant IRC~+10420. As mentioned earlier, the identification of emission  in the LMC YSGs is limited by the lack of appropriate spectra. 
We find only a few  with emission lines based on published notes. For example LHa 120-S159 (F2 I) has weak Hydrogen emission most likely due to nebulosity \citep{Henize}, but emission present in the H$\alpha$ absorption lines in Echelle spectra of HD 271182 and HD 269953 \citep{Kour} is attributed to atmospheric activity. 

\vspace{2mm} 

Sk -69 147, classified as F5 Ia (ARDB 246) by
\citet{Ardeberg}, and F5 I by \citet{Sand} is identified on SIMBAD with MWC 112 which is sometimes listed as an LBV or LBV candidate.  MWC 112 is described as Beq with P Cyg type emission in the MWC Catalog \citep{MWC}.   \citet{Ardeberg}  often give notes about the spectra including emission lines, but there is no mention of emission for Sk -69 147.  Thus this star appears to be a normal F supergiant. \citet{vanGenderen}, however, identify MWC 112 with HD 269582 = MWC 112= HV 5495=Sk -69 142a, a WR star, WN10h \citep{Crowther}, and \citep{Sand} lists -69 142a as a Be star. Obviously the WR star is the better candidate for MWC 112. We suggest that the YSG Sk -69 147 is not MWC 112.

\vspace{2mm}

The A-type FYPS, HD 269661 (A0Ia0e) has Hydrogen emission and suspected He I emission \citep{Ardeberg}. Its SED also shows a small excess due to dust. A second FYPS  HD 269781 also has Hydrogen emission but no circumstellar dust. In addition to being pulsators, both stars may have winds and enhanced mass loss due to their post-RSG  state. 

\section{Comments on the Evolutionary State}

The nine stars in Table 1 are shown on an HR Diagram for evolved supergiants in the LMC in Figure 2. Their high lumnosity has already been noted. These stars are all above the 20 M$_{\odot}$ track at about 10$^{5}$ L$_{\odot}$ and the nominal upper mass limit to the progenitors of the Type IIP supernovae. These stars are about 1/3 of the YSGs above  10$^{5}$ L$_{\odot}$, similar to the fraction of post-RSG candidates we found in M31 and M33 \citep{Gordon}.  

\vspace{2mm}

The fate of RSGs with initial masses above 20 M$_{\odot}$ is debated. {For example, in a recent paper \citet{Pedersen} argued that the FYPS were the result of contamination in the TESS data, but \citet{Dorn22} identified likely contaminated stars and removed them from their sample.}      
 Six of the nine stars have circumstellar dust which distinguishes  them from the other YSGs.  {The two FYPS with dusty ejecta have the largest circumstellar excesses and highest mass loss rates including the star with the detached shell from a prior RSG stage.}  We conclude  that with their dusty ejecta and continued mass loss, these yellow hypergiants are   
 candidates for post-RSG evolution among the yellow and red supergiants in the LMC.  

\begin{figure}
\plotone{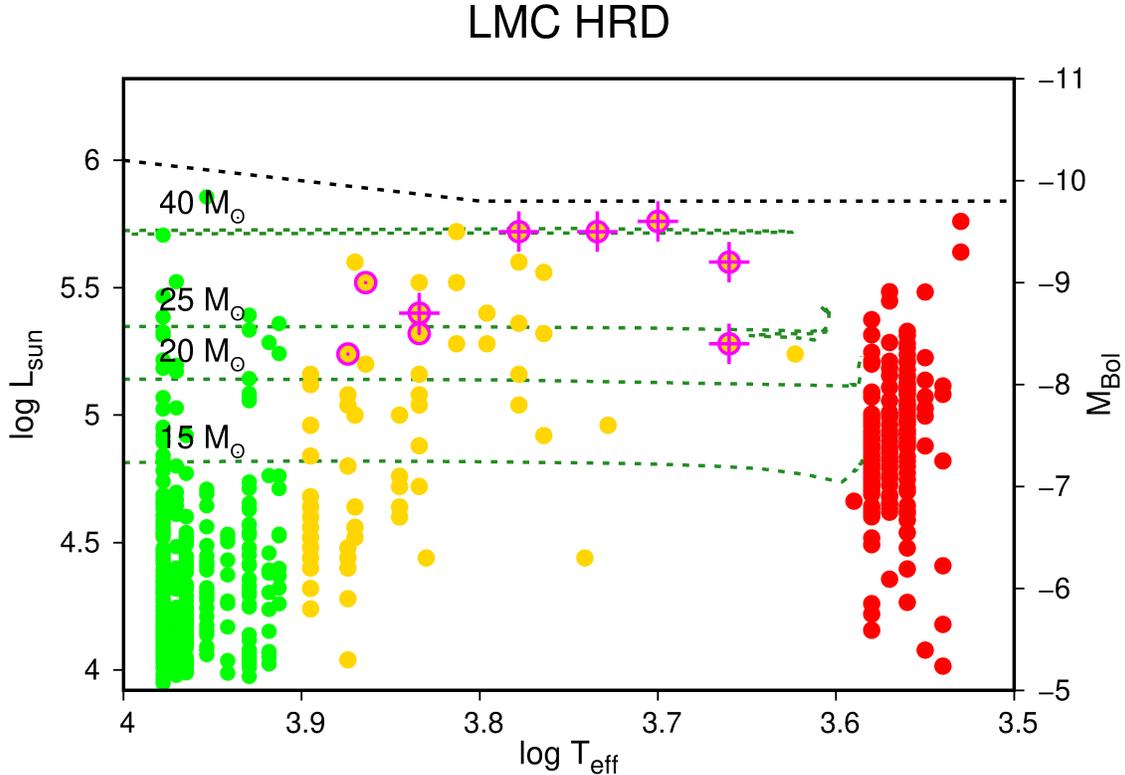}
	\caption{The upper HR Diagram for the evolved warm and cool supergiants, Log T$_{eff}$ 4.0 to 3.5. The A-type supergiants are shown in green, the YSGs in yellow, and the RSGs in red.  The FYPS stars from \citep{Dorn22} and the four additional yellow supergiants with circumstellar dust are plotted with open circles, and the six stars with circumstellar dust are highlighted with a cross. The dashed line at the top is the Humphreys-Davidson limit from previous work.  The evolutionary tracks are from \citep{Eggen}. }
\end{figure}

\end{document}